\documentclass[conference]{IEEEtran}
\IEEEoverridecommandlockouts
\usepackage{cite}
\usepackage{amsmath,amssymb,amsfonts}
\usepackage{algorithmic}
\usepackage{graphicx}
\usepackage{textcomp}
\usepackage{xcolor}
\def\BibTeX{{\rm B\kern-.05em{\sc i\kern-.025em b}\kern-.08em
    T\kern-.1667em\lower.7ex\hbox{E}\kern-.125emX}}
\begin{document}
\pagestyle{plain}

\title{Bilateral Trade Flow Prediction by Gravity-informed Graph Auto-encoder\\
}

\author{\IEEEauthorblockN{Naoto Minakawa}
\IEEEauthorblockA{\textit{Department of Systems Innovation}\\  \textit{School of Engineering}\\
\textit{The University of Tokyo}\\
Tokyo, Japan \\
naoto-minakawa@g.ecc.u-tokyo.ac.jp}
\and
\IEEEauthorblockN{Kiyoshi Izumi}
\IEEEauthorblockA{\textit{Department of Systems Innovation}\\  \textit{School of Engineering}\\
\textit{The University of Tokyo}\\
Tokyo, Japan \\
izumi@sys.t.u-tokyo.ac.jp}
\and
\IEEEauthorblockN{Hiroki Sakaji}
\IEEEauthorblockA{\textit{Department of Systems Innovation}\\  \textit{School of Engineering}\\
\textit{The University of Tokyo}\\
Tokyo, Japan \\
sakaji@sys.t.u-tokyo.ac.jp}
}

\IEEEoverridecommandlockouts
\IEEEpubid{\makebox[\columnwidth]{978-1-6654-8045-1/22/\$31.00~\copyright2022 IEEE \hfill}
\hspace{\columnsep}\makebox[\columnwidth]{ }}

\maketitle

\begin{abstract}
The gravity models has been studied to analyze interaction between two objects such as trade amount between a pair of countries, human migration between a pair of countries and traffic flow between two cities. Particularly in the international trade, predicting trade amount is instrumental to industry and government in business decision making and determining economic policies. Whereas the gravity models well captures such interaction between objects, the model simplifies the interaction to extract essential relationships or needs handcrafted features to drive the models. Recent studies indicate the connection between graph neural networks (GNNs) and the gravity models in international trade. However, to our best knowledge, hardly any previous studies in the this domain directly predicts trade amount by GNNs. We propose GGAE (Gravity-informed Graph Auto-encoder) and its surrogate model, which is inspired by the gravity model, showing trade amount prediction by the gravity model can be formulated as an edge weight prediction problem in GNNs and solved by GGAE and its surrogate model. Furthermore, we conducted experiments to indicate GGAE with GNNs can improve trade amount prediction compared to the traditional gravity model by considering complex relationships.
\end{abstract}


\section{Introduction}
International trade becomes increasingly important these days. According to United Nations Conference on Trade and Development (UNCTAD), global trade continues to grow and reached a record high in 2021, approximately 28.5 trillion dollars, which is an increase of almost 13\% relative to pre-pandemic levels \cite{UNCTAD2022}. On the other hand, wider variety of data becomes available regarding international trade. For example, The World Trade Organization (WTO) provides data related to international trade such as trade flows, tariffs, non-tariff measures (NTMs) and trade in value \footnote[1]{https://www.wto.org/english/res\_e/statis\_e/statis\_e.htm}. Not only structured datasets such as tabular data, but also unstructured data such as report is available.

To analyze international trade, the gravity models have been actively studied by many economists \cite{Tinbergen1962, EatonKortum2002, AndersonWincoop2003, Anderson1979}. The gravity model of international trade \cite{Tinbergen1962} is a model inspired by Isaac Newton’s universal law of gravitation which predicts bilateral trade flows based on the economic dimensions and distance between two countries. For example, GDP is adopted as economic dimensions. As mentioned later, the model states that trade flow between a pair of countries is proportional to the multiplication of economic dimensions of the countries divided by the distance between the countries.

While traditional and newer gravity models well captures the essence of international trade \cite{Tinbergen1962, EatonKortum2002, AndersonWincoop2003, Anderson1979}, they often cannot deal with complex relationships between countries on top of large volume of structured and unstructured data at the same time. In this regards, neural networks (NN) are suitable for handling such fusion of structured and unstructured data. Particularly, graph neural networks (GNNs) considers network topology, which can be considered the interaction or relationships of countries in the context of international trade, on top of data fusion. Therefore, GNNs are well-suited to handle complex relation between trading countries on top of wide variety and larger volume of structured and unstructured trade related- data which becomes available.

Indeed, recent studies indicate the connection between GNNs and the gravity models in international trade \cite{Panford-Quainoo20BilateralGNN, Monken_2021, Verstyuk22Gravity}. However, to our best knowledge, hardly any previous studies in the this domain directly predicts trade amount by GNNs, since standard link prediction task in the context of GNNs is to predict the link between nodes (i.e. trade existence between a pair of countries). 

In this paper, we focus on how to formulate trade amount prediction in GNNs, since it becomes possible to predict  trade amount by considering relationships between countries on top of fusion of structured and unstructured data once the problem is formulated in the GNN framework. In order to predict trade amount in GNNs, we propose GGAE (Gravity-informed Graph Auto-encoder), which is inspired by a the gravity model, showing the gravity model can be formulated as edge weight prediction problem in GNNs and solved by GGAE. Furthermore, we conducted experiments to indicate use of GNNs to consider more complex relationships among countries can improve trade prediction compared to the traditional gravity model.

The main contributions of this paper are summarized as follows:

\begin{itemize}
 \item We proposed a novel graph auto-encoder called Gravity-informed Graph Auto-encoder (GGAE) and its surrogate model, which is inspired by the gravity model and deemed to work well in the context where the gravity model is known to hold true.
 \item We showed trade amount prediction by the traditional gravity model \cite{Tinbergen1962} is formulated as edge weight prediction task in a GNN and trade amount is predicted by using GGAE and its surrogate model.
 \item We conducted experiment to show application of GGAEs to international trade amount prediction can take wider range of trade relationships into account compared to traditional methods, resulting in better prediction results.
\end{itemize}

The remainder of the paper is organized as follows. Section II introduces the related work on prediction of international trade amount. Section III introduces preliminary concepts. Section IV states proposed methods for bilateral trade flow prediction. Section V states about the evaluation including dataset, task and experiment. Section VI discusses the results. Finally, Section VII presents the conclusion.

\section{Related Work}
The gravity models have been studied to analyze interaction between two objects. Other than international trade, the gravity models are applied to predict human migration between a pair of countries \cite{Karemara2000GravityMigration, Lewer2008GravityImmigration} and traffic flow between two cities \cite{Ru2010GravityTraffic, Sayed2017GravityTraffic}, for instance. 

Particularly in the context of international trade, predicting trade amount is an important topic since it is instrumental to industry and government in determining economic policies. Many economists have been studying the gravity models of trade to describe bilateral trade flows between a pair of countries \cite{Tinbergen1962, EatonKortum2002, AndersonWincoop2003, Anderson1979}. Since there are so many researches on the gravity models in international trade that it is difficult to refer to all of them, please refer to more detailed reviews \cite{HeadMayer2014}.

Different approaches have also been taken to predict the trade amount. Recently, machine learning and neural network approaches have been popular and taken to tackle with the trade flow prediction problem. Wohl et al.\cite{Wohl2018NNIntlTrade} used shallow NNs to improve international trade forecasting. Huang et al. \cite{Huang2022IntlTradeGNN} applied deep neural networks (DNN) to international trade quantification, and conducted comparative analysis of various prediction methods including ARIMA, shallow NNs, and DNNs. Kottou et al. \cite{Kottou2020wavelets} proposed an algorithm based on Machine Learning methods combined with Wavelet Transforms to predict the bilateral trade flow between a pair of countries by using economic indicators as input data. 

Recent studies indicates the connection between GNNs and the gravity models in international trade. Representation of the international trade as a graph or network is considered effective since we can capture complex topological relation between countries.

For example, Panford-Quainoo et al. \cite{Panford-Quainoo20BilateralGNN} studies connection between a trade gravity model and GNNs. They formulate a country GDP classification task as a node classification problem in GNNs and a trade partner finding task as a link prediction problem in GNNs, however, they have not conducted trade amount prediction, which can be interpreted as an edge weight prediction problem in GNNs. Monken et al. \cite{Monken_2021} suggest a methodology by applying GNNs to analyze the time-varying structure of the network of bilateral country trade, modeling causality in international trade to predict future patterns under unforeseen circumstances. Verstyuk et al. \cite{Verstyuk22Gravity} discuss that GNNs as a natural and theoretically appealing class of models for international trade, demonstrating the theoretical connection and empirical results fitted to a large panel of annual-frequency country-level data to analyze bilateral accessibility. 

Whereas some previous works studies connection between GNNs and trade gravity models, to our best knowledge, trade amount prediction has not been conducted, while it is an important task. Presumably, this is due to that the standard link prediction problem in GNNs is to predict the existence of an edge between two nodes, not to predict the amount associated with the edge. While a link between a pair of nodes is important in all the contexts such as chemistry (i.e. representing protein as a graph), financial network, and social networks, weight information on the edge is available and important with the limited type of networks. In the financial networks, edge weight information between two nodes such as trade amount between countries, debt between banks are often more important than the link itself.

As far as financial networks are concerned, there is only one previous work found which aims to predict transaction amount in Bitcoin network \cite{Sharma2020Bitcoin}. They used the Temporal Graph Convolutional Network (T-GCN) to first extract the topological features from the transaction data then extract the temporal features at the current timestamp along with information of previous timestamps using Gated Recurrent Unit (GRU) in order to predict the output at the next timestamp. However, the decoder used is simply a GRU and any edge information is not taken into account unlike our approach. On the other hand, link prediction problems have been well studied. For the link prediction problem in GNNs, graph auto-encoder (GAE) is one of the most well-known and simplest approach \cite{Kipf2016GAE}. Regarding the applications to financial networks, there are several works to apply link prediction problems to predict existence of trades in financial networks such as banking transaction networks \cite{Fujitsuka2019, Shumovskaia2021, Minakawa2022}.

Our study focus on the interpretation of a gravity model as a modified version of GAE \cite{Kipf2016GAE} to apply to the edge weight prediction problem, suggesting the potential of applying GGAEs to international trade amount prediction by considering more complex relationships; not only the neighboring counterparties but also indirect counterparties, which are multiple hops away from the country.

\section{Preliminaries}

In this section, we present the preliminary concepts which are used as foundations for our methodology.

\subsection{The Gravity Models in International Trade}

It was found that trade flow between a pair of countries is expressed with the following equation which resembles Isaac Newton’s universal law of gravitation \cite{Tinbergen1962}.

\begin{align}
\text{Trade Flow}_{u,v} \approx \gamma \frac{GDP_u GDP_v}{distance(u,v)}
\end{align}

where $GDP_u$ and $GDP_v$ represents GDP of countries $u,v$, $distance(u,v)$ represents the distance between countries $u$ and $v$, and $\gamma$ represents a scaler value.

The gravity model represents that trade flow between countries $u$ and $v$ is proportional to the product of GDP of countries $u$ and $v$ divided by the distance between countries $u$ and $v$. The magnitude of this trade flow increases when GDP of countries $u,v$ increases and decreases when the distance between countries $u$ and $v$ increases.

In a logarithmic form, the equation (1) is equivalent to the following equation:

\begin{align}
log(\text{Trade Flow}_{u,v}) & \approx
log(\gamma) + log(GDP_u) + log(GDP_v) \notag \\
& - log(distance(u,v))
\end{align}

\subsection{Graph Convolutional Networks}
GCN \cite{Kipf2017GCN} is one of the most popular models among GNNs. The model structure of GCN is relatively simple and often shows good performance. Therefore, we used GCN to obtain node embeddings.

Let us denote a network as $G(V, E)$, node as $u \in V$, edge as $(u,v) \in E$, neighbor nodes of node $u \in V$ as $\mathcal{N}(u)$, $\textbf{A} \in \mathbb{R}^{|V|\times |V|}$ as the adjacency matrix, $\tilde{\textbf{A}}$ as the adjacency matrix with added self-connections, $\tilde{\textbf{D}}$ is a degree matrix \footnote[2]{Degree matrix is represented as $\tilde{\textbf{D}}_{i,i} = \sum_{j} \tilde{\textbf{A}}_{i,j}.$}. We denote $\textbf{H}_u^{(k)}$ as an embedding of node $u$ in $k$-th layer in the neural network. $\textbf{H}_u^{(0)}$ is equivalent to the feature vector of node $u$, which is a look up of node feature matrix $\textbf{X} \in \mathbb{R}^{|V|\times d}$. $W^{(k)} \in \mathbb{R}^{|\text{Hidden dim.}|\times |\text{Input dim.}|}$ is a shared trainable parameter in $k$-th layer in the network, where $|\text{Hidden dim.}|$ and $|\text{Input dim.}|$ represents the dimensions of the hidden and input layers, respectively. $\sigma$ is an activation function. We used $\textbf{ReLU}$ which is widely adopted. Then, GCN \cite{Kipf2017GCN} is formulated as follows:

\begin{align}
 \textbf{H}^{(k+1)}=\sigma\left(\tilde{\textbf{D}}^{-1 / 2} \tilde{\textbf{A}} \tilde{\textbf{D}}^{-1 / 2} \textbf{H}^{(k)} \textbf{W}^{(k)}\right).
\end{align}

\subsection{Graph Auto-encoder (GAE)}
Graph auto-encoder (GAE) \cite{Kipf2016GAE} is one of the most well-known and simplest methods used for link prediction tasks in GNNs. GAE predicts link between a pair of nodes by reconstructing the original adjacency matrix using node embeddings. Once node features are passed through GNNs, we obtain embedded node features $\textbf{H} = \textbf{H}^{(k)}$. Then, the obtained embedding is decoded using a GAE \cite{Kipf2016GAE}. With GAE, for a particular edge, its linkage is reconstructed as follows\footnote[3]{The adjacency matrix $\textbf{A}$ is reconstructed as $\textbf{A} \approx \textbf{sigmoid}(\textbf{H} \textbf{H}^{T})$}:

\begin{align}
 \textbf{A}_{u,v} \approx \textbf{sigmoid}(\textbf{H}_u \textbf{H}^{T}_v).
\end{align}

Please note that we can predict if there is a trade between a pair of countries (in another words to find a trade partner) by applying GAE to international trade network as studied in previous work \cite{Panford-Quainoo20BilateralGNN}. However, we cannot to predict transaction amount between the a pair of countries by simply applying GAE.

\section{Methodology}

\subsection{Gravity-informed Graph Auto-encoder}
We modify the aforementioned GAE \cite{Kipf2016GAE} into gravity-informed way to predict not linkage but edge weight, since GAE is one of the most well-known and simplest methods used for link prediction as aforementioned. It is also clear to see how it can be changed to gravity-informed fashion.

Let us denote the trade amount matrix of the original network as $\textbf{A}^{amt}$, which contains all the trade amounts between two nodes. That is to say, trade flow between $u$ and $v$ is $\textbf{A}^{amt}_{u,v}$. We also denote \textbf{W} as a weight vector, \textbf{B} as a bias vector, $\textbf{E} \in \mathbb{R}^{|V|\times |V|}$ as the edge feature matrix, which contains all the distances between two countries in our case. In the simplest form, GGAE aims to reconstruct the trade amount matrix $\textbf{A}^{amt} \in \mathbb{R}^{|V|\times |V|}$ as follows:

\begin{align}
\textbf{A}^{amt}_{u,v} \approx \textbf{W} \textbf{H}^e_{u,v} + \textbf{B}
\end{align}

where, 

\begin{align}
\textbf{H}^e_{u,v} = \textbf{H}_u \textbf{H}^T_v \textbf{H}_{u,v}^{-1}.
\end{align}

We suggested the simplest form as a starting point, however, it can be more expressive. For instance, we can consider

\begin{align}
\textbf{A}^{amt}_{u,v} \approx \textbf{MLP}(\textbf{H}^e_{u,v}).
\end{align}

Otherwise, the following form can be considered as a surrogate model for the equation (8). Whereas the equation (8) specifies the gravity model- relation explicitly, the surrogate model aims to learn the relationship. This is deemed to be effective when we use different kinds of node features where not all node features necessarily have the gravity model- relationships. It is also effective when we want to consider multi-dimensional edge features.

\begin{align}
\textbf{A}^{amt}_{u,v} \approx \textbf{MLP}(\textbf{H}_u ||\textbf{H}_v||\textbf{E}_{u,v})
\end{align}

where $||$ represents concatenation of vectors.

Both GGAE and the surrogate decoder of GGAE consider edge features on top of node features, enabling decoder function to have more representational power than the original GAE.

\subsection{Connection with the gravity model}

In the most general form, GNNs are represented as the following equation with some permutation-invariant \footnote[4]{A function $f: X(\Omega) \rightarrow Y$ is $\mathfrak{G}-$invariant if $f(\rho(\mathfrak{g})x)=f(x)$ for all $\mathfrak{g} \in \mathfrak{G}$ and $x \in X(\Omega)$, i.e., its output is unaffected by the group action on the input. \cite{Bronstein2021Geometric}} function $\oplus$ and some permutation-equivariant\footnote[5]{A function $f: X(\Omega) \rightarrow X(\Omega)$ is $\mathfrak{G}-$equivariant if $f(\rho(\mathfrak{g}))=\rho(\mathfrak{g})f(x)$ for all $\mathfrak{g} \in \mathfrak{G}$, i.e., group action on the input affects the output in the same way. \cite{Bronstein2021Geometric}} function $\psi$ \cite{Bronstein2021Geometric}.

\begin{align}
\textbf{H}^{(k)}_u = \phi(\textbf{H}^{(k-1)}_u, \underset{v \in \mathcal{N}(u)}{\displaystyle \oplus} \psi(\textbf{H}^{(k-1)}_u, \textbf{H}^{(k-1)}_v))
\end{align}

An identity mapping trivially satisfies permutation equivariance and permutation invariance. In that sense, if we do not apply an activate function, applying identity mapping to the original node features can be considered as applying GNNs. 
Specifically, when we use an identity mapping instead of GCN,  $\textbf{H}^0_u = \textbf{X}$ is an one-dimensional feature (i.e. GDP of countries) and $\textbf{E}_{u,v}^{-1}$ is an one-dimensional feature (i.e. distance between a pair of countries), $\textbf{H}_u \textbf{H}^T_v \textbf{E}_{u,v}^{-1}$ in equation (6) exactly represents the gravity model.

\begin{align}
\text{Trade Flow}_{u,v} = \textbf{A}^{amt}_{u,v} = \textbf{H}_u \textbf{H}^T_v \textbf{E}_{u,v}^{-1} = \frac{GDP_u GDP_v}{distance(u,v)}.
\end{align}

\section{Evaluation}
In the previous section, we showed trade amount prediction by the traditional gravity model \cite{Tinbergen1962} can be formulated as edge weight prediction task with GNNs and trade amount is predicted by using GGAE or its surrogate model. In this section, we conduct experiment to show application of GGAEs to international trade amount prediction can take wider range of trade relationships into account compared to traditional methods, resulting in better prediction results.

\subsection{Dataset}
The data used for the experiment were taken from The CEPII Gravity Database \footnote[6]{http://www.cepii.fr/CEPII/en/bdd\_modele/bdd\_modele\_item.asp?id=8}. The CEPII is the leading French research center for the world economy founded in 1978, being part of the network coordinated by France Strategy, within the Prime Minister's services. The CEPII produces databases and provides a platform for debate among academics, experts, practitioners, decision makers and other private and public stakeholders. The CEPII Gravity database gathers a set of variables, such as year, exporter, importer, trade flows as well as geographic, cultural, trade facilitation and macroeconomic variables, for the determinants of international trade in a single place. Data is obtained from many different sources such as the World Bank, the WTO and the IMF.

In our experiment, we used such data attributes as exporter, importer, exporter-importer distance, exporter GDP, importer GDP, export values, and import values. Records are removed where GDP and distance are missing in order to define the gravity model- relationship properly (i.e. the right hand side of the equation (1) becomes zero when GDP is missing; the right hand side of the equation (1) goes to infinity when distance is missing). The reason we limit the scope of features is to guarantee that the gravity model- relation hold true among selected features.

As shown in Fig.1, it is confirmed that trade amount is proportional to the gravity model- relation indicated in the equation (1); multiplication of GDPs of a pair of countries divided by the distance between them. It is appropriate to plot the trade amount in the logarithmic scale, since its distribution is highly skewed. So are the GDPs. In order to better predict trade amount, we converted these numeric features to logarithmic scale. 

\begin{figure}[htbp]
\centerline{\includegraphics[width=\linewidth]{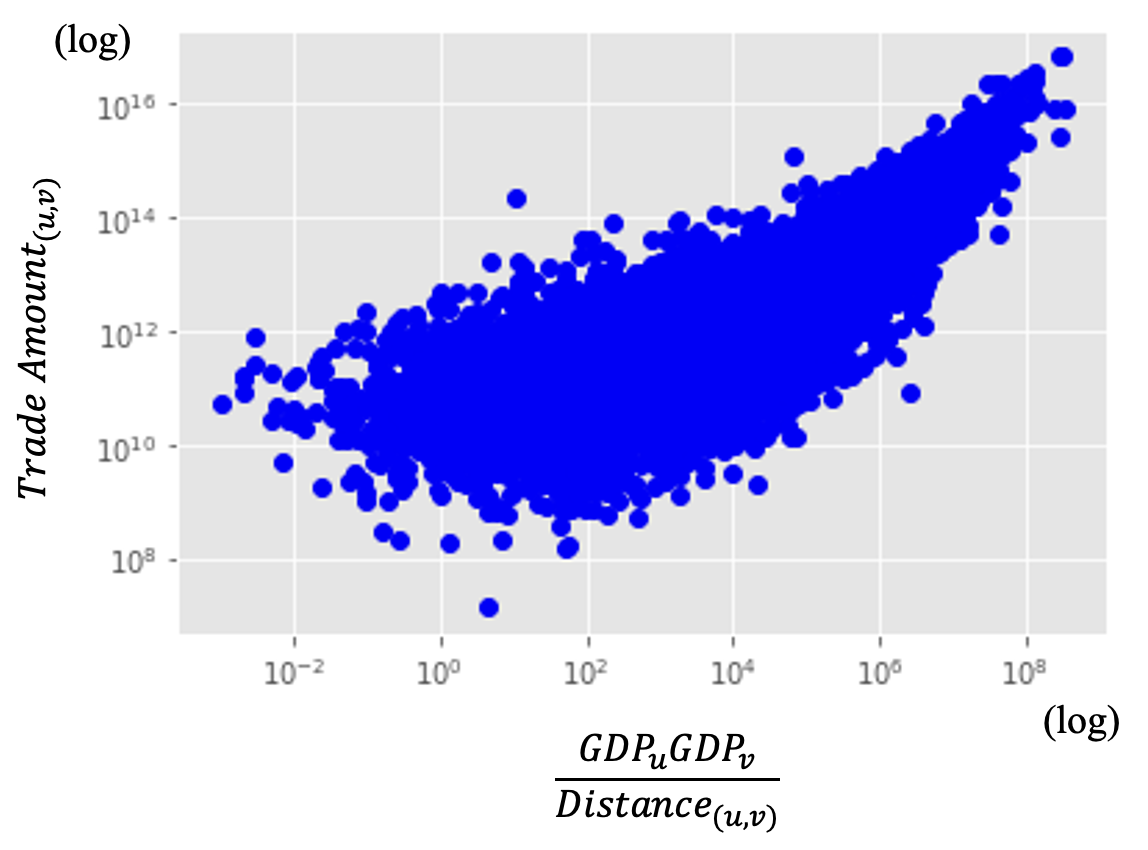}}
\caption{Log-log plot of trade amount and the gravity model- relationships between a pair of countries}
\label{Fig1}
\end{figure}

Constructed international trade network is as illustrated in Fig.2. The network includes 186 countries and 13811 trade transactions. GDPs of countries are given as node features, trade amount and distance between a pair of countries are given as edge features.

\begin{figure}[htbp]
\centerline{\includegraphics[width=\linewidth]{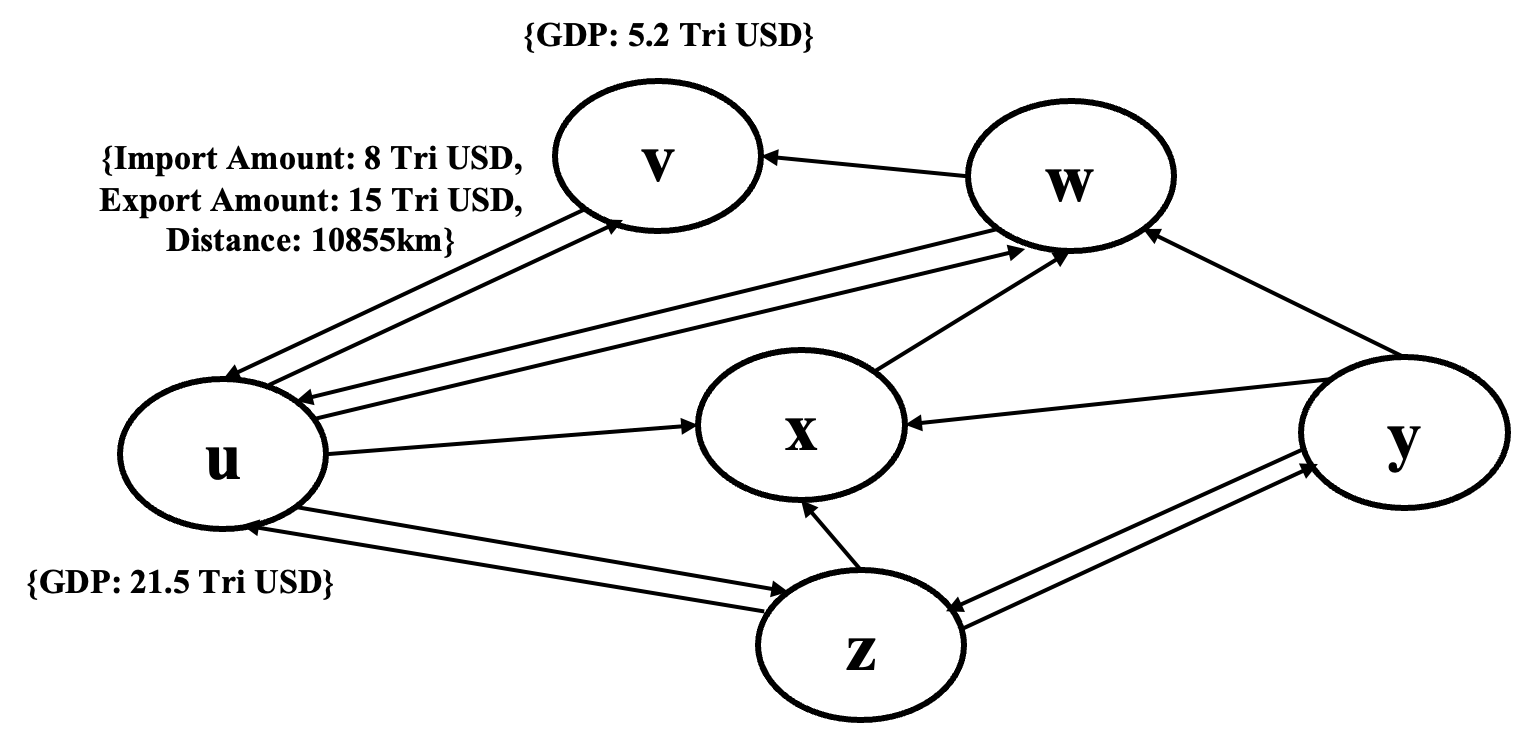}}
\caption{International trade network (an illustrative example)}
\label{Fig2}
\end{figure}

\begin{table*}[ht]
\caption{Experiment}
\begin{center}
\begin{tabular}{|c|c|c|c|c|c|c|c|}
\hline
pattern & 
\begin{tabular}{c}
node feature \\ $\textbf{X}_u = \textbf{H}^0_u$
\end{tabular} & 
\begin{tabular}{c}
edge feature \\ $\textbf{E}_{u,v}$
\end{tabular} & 
\begin{tabular}{c}
encoder \\ $f^E(\textbf{H}_u)$
\end{tabular} & 
\begin{tabular}{c}
decoder \\ $f^D(\textbf{H}_u, \textbf{H}_v)$
\end{tabular} & 
\begin{tabular}{c}
RMSE \\ (average)
\end{tabular} &
\begin{tabular}{c}
RMSE \\ (max)
\end{tabular} &
\begin{tabular}{c}
RMSE \\ (min)
\end{tabular} \\
\hline \hline
(1) & $log(GDP_u)$ & $log(dist(u,v))$ & 
\begin{tabular}{c}
$f^E(\textbf{H}_u) = \textbf{H}_u$ \\ (Identity function)
\end{tabular} &
\begin{tabular}{c}
$W \textbf{H}^e_{u,v} + B$ \\ $\textbf{H}^e_{u,v} = \textbf{H}_u + \textbf{H}_v - \textbf{E}_{u,v}$ \\
$= log(\frac{\textbf{H}_u \textbf{H}_v}{dist(u,v)})$
\end{tabular} &
5.137 &
5.650 &
4.967 \\
\hline
(2) & $ log(GDP_u)$ & $log(dist(u,v))$ &
\begin{tabular}{c}
$f^E(\textbf{H}_u)$ \\ $ = GCN^{(1)}(\textbf{H}_u)$
\end{tabular} &
\begin{tabular}{c}
$W \textbf{H}^e_{u,v} + B$ \\ $\textbf{H}^e_{u,v} = \textbf{H}_u + \textbf{H}_v - \textbf{E}_{u,v}$ \\
$= log(\frac{\textbf{H}_u \textbf{H}_v}{dist(u,v)})$
\end{tabular} &
5.017 &
5.162 &
4.795 \\
\hline
(3) & $ log(GDP_u)$ & $log(dist(u,v))$ & \begin{tabular}{c}
$f^E(\textbf{H}_u)$ \\ $ = GCN^{(1)}(\textbf{H}_u)$
\end{tabular} &
$MLP([\textbf{H}_u||\textbf{H}_v||\textbf{E}_{u,v}])$ &
4.706 &
4.889 &
4.548 \\
\hline
(4) & $ log(GDP_u)$ & $log(dist(u,v))$ & \begin{tabular}{c}
$f^E(\textbf{H}_u)$ \\ $ = GCN^{(2)}(\textbf{H}_u)$
\end{tabular} &
\begin{tabular}{c}
$W \textbf{H}^e_{u,v} + B$ \\ $\textbf{H}^e_{u,v} = \textbf{H}_u + \textbf{H}_v - \textbf{E}_{u,v}$ \\
$= log(\frac{\textbf{H}_u \textbf{H}_v}{dist(u,v)})$
\end{tabular} &
4.602 &
4.710 &
4.428 \\
\hline
(5) & $ log(GDP_u)$ & $log(dist(u,v))$ & \begin{tabular}{c}
$f^E(\textbf{H}_u)$ \\ $ = GCN^{(2)}(\textbf{H}_u)$
\end{tabular} &
$MLP([\textbf{H}_u||\textbf{H}_v||\textbf{E}_{u,v}])$ &
\textbf{4.122} &
\textbf{4.377} &
\textbf{3.879} \\
\hline
\end{tabular}
\label{tab1}
\end{center}
\end{table*}

\subsection{Task and evaluation metrics}
Link prediction is one of three typical tasks in GNNs. It is binary prediction of the existence of an edge between a pair of nodes. One popular loss function used in link prediction task is the cross-entropy loss with negative sampling \cite{Hamilton2020GRL}. ROC-AUC score is often used as evaluation metrics in link prediction. 

In our case, however, it is not usual binary prediction of an edge between a pair of nodes. Instead, it is to predict an edge weight between a pair of nodes, resulting in an edge-level regression task. Therefore, we adopted MSE as loss function as well as RMSE as an evaluation metrics, which is one of the most popular combination for regression tasks.

\subsection{Experiment}
We conducted experiments to confirm if use of GGAE with GNNs can improve traditional trade amount prediction compared to the gravity model by considering complex relationships. We formulate the traditional gravity model as a naive GNN as stated in the section IV.B, comparing with the deeper GNNs with different decoder functions. Specifically, we compared five patterns: (1) the gravity model, (2) 1 layer GCN and GGAE, (3) 1 layer GCN and MLP (the surrogate decoder of GGAE), (4) 2 layer GCN and GGAE, (5) 2 layer GCN and MLP (the surrogate decoder of GGAE).

For the implementation, we used dgl \footnote[7]{deep graph library; https://www.dgl.ai/} for implementation of GNNs and scikit-learn \footnote[8]{https://scikit-learn.org/stable/} for computing evaluation metrics. 
We adopted MSE as loss function and RMSE as evaluation metrics as stated in the previous subsection, 0.01 as learning rate, 1000 as the number of epochs, Adam as an optimizer. For GGAE, we adopted the functional defined in the equation (8). For the surrogate decoder of GGAE, we adopted 3 layer MLP as MLP function defined in the equation (11). We use 66\% of the entire edges as train edges, 33\% of the entire edges as test edges to compare with different methods. We run the same experiments 10 times for each method. Then, we compare average, max and minimum of RMSE scores for each method. Those results are presented in Table 1.

\section{Discussion}
As presented in Table 1, patterns (2)-(5) outperforms (1) when we compare 5 different models. This is due to that GCNs can take more complex relationships into account; 1 layer GNNs considers the neighboring counterparties and 2 layer GNNs considers neighboring counterparties which are 2 hops away from the specific country.

Comparing 1 layer GCNs (2)-(3) and 2 layer GCNs (4)-(5), 2 layer GCNs shows better performance. This is considered due to that 2 layer GCNs consider neighboring countries which are further away from the specific countries, whereas 1 layer GCNs consider only the neighboring countries. 

Comparing GGAE (2),(4) and MLP decoder (3),(5) as its surrogate decoder, the surrogate decoders show slightly better performance. One potential reason is that we adopted 3 layer MLP as the functional used for the decoder, enabling learning relationships more flexibly from data than the original GGAE which explicitly specifies the gravity model- relationship. It is possible to see more distinct differences between GGAE and MLP decoder, when we conduct experiment with more variety of node and edge features, because not all features necessarily satisfy the gravity model- relationships. While GDP and distance are known to satisfy the gravity model- relationships, there are no guarantee that other node and edge features satisfy the relationships. For instance, if we adopt the embedded vectors of textual data about the culture of countries as one of node features and distances between countries as edge features, it does not necessarily satisfy the relationship. Our assumption is that when node features and edge features satisfy the gravity model- relationships clearly, GGAE should perform well. If that is not the case, the surrogate MLP decoder should perform well as it learn relationships from actual data more flexibly. In summary, the key of our work is to show decoder can be made gravity-informed and trained end-to-end rather than naively apply various supervised learning models to obtained embeddings for trade amount prediction.

As a future work, we will first retrieve information from other data sources to study if the findings of this paper still hold when the missing records were added back. Then, we will explore more variety of node and edge features to verify if GGAE can further improve trade amount prediction. More specifically, we will consider structured and unstructured data such as textual information about countries as node features. Also, we will study the detailed behavior of GGAE and the MLP decoder as its surrogate when we increase the complexity of node and edge features. Finally, we will consider to apply GGAEs to different kinds of network datasets such as other financial networks like banking transaction networks and inter-firm networks, traffic flow networks and human migration networks where the gravity models are known to hold true.

\section{Conclusion}
In this paper, we proposed a novel approach for bilateral trade flow prediction by GNNs which can scale up to more complex datasets. Specifically, we introduced a novel graph auto-encoder called Gravity-informed Graph Auto-encoder (GGAE) and its surrogate model, which is inspired by the gravity model. We showed the traditional gravity model \cite{Tinbergen1962} can be formulated as edge weight prediction problem in GNNs and trade amount can be predicted by GGAE and its surrogate model. Furthermore, we conducted experiment to show application of GGAE to international trade amount prediction can take wider range of topological information into account compared to traditional methods, resulting in better prediction results. As a future work, we will first study if the results still hold when the missing records were added back. Then, we will explore more variety of node and edge features. Also, we will study the detailed behavior of GGAE and its surrogate model when we increase the complexity of node and edge features. Finally, we will consider to apply GGAEs to different kinds of network datasets such as other financial networks, traffic flow networks and human migration networks where the gravity models hold true.

\section*{Acknowledgement}
This work was supported by JST-Mirai Program Grant Number JPMJMI20B1, Japan.

\bibliographystyle{IEEEtran}
\bibliography{main}


\end{document}